\documentclass[prb,aps,twocolumn,showpacs,floatfix]{revtex4}
\usepackage{graphicx}

\begin{document}
\title{Theory of transverse spin dynamics in a polarized Fermi liquid
and an itinerant ferromagnet} \author{V. P. Mineev }
\affiliation{Commissariat \'a l'Energie Atomique, DSM/DRFMC/SPSMS
38054 Grenoble, France} \date{May 1, 2005}

\begin{abstract}
The linear equations for transverse spin dynamics in a weakly polarized
degenerate Fermi liquid with arbitrary relationship between
temperature $T$ and polarization $\gamma H$ are derived from
Landau-Silin phenomenological kinetic equation with general form of
two-particle collision integral.  Unlike the previous treatment
where Fermi velocity  and density of states have been taken as constants
independent of polarization here we made derivation free from this
assumption.  The obtained equations are applicable for description of
spin dynamics in paramagnetic Fermi liquid with finite polarization as
well in an itinerant ferromagnet.  In both cases transverse spin wave
frequency is found to be proportional to $k^{2}¥$ with complex
constant of proportionality (diffusion coefficient) such that the
damping has a finite value at $T=0$.  The polarization dependence of
the diffusion coefficient is found to be different for a polarized
Fermi liquid and for an itinerant ferromagnet.  These conclusions are
confirmed by derivation of transverse spin wave dispersion law in
frame of field theoretical methods from the integral equation for the
vortex function.  It is shown that similar derivation taking into
consideration the divergency of static transverse susceptibility also
leads to the same attenuating spin wave spectrum.

\end{abstract}

\maketitle

\section{Introduction}

Since it appearence in the pioneering papers of V.P.Silin \cite{Sil}
and A.J.Leggett \cite{Leg} the theory of spin dynamics in spin
polarized Fermi liquid has a long history mainly concerned with the
problem of the zero-temperature transverse spin-wave attenuation.  The
calculations of transverse spin-diffusion coefficient in dilute
degenerate Fermi gas with arbitrary polarization was done for the
first time in the papers by W.Jeon and W.Mullin \cite{JM} where the
low temperature saturation of corresponding relaxation time has been
established.  About the same time A.Meyerovich and K.Musaelyan
\cite{Me,Mu} have derived the spin kinetics in the polarized Fermi
liquid from microscopic theory and also came to the same conclusion. 
A derivation and an exact solution of the kinetic equation in the
$s$-wave scattering approximation for dilute degenerate Fermi gas with
arbitrary polarization at $T=0$ and for a small polarization at $T\ne
0$ have been obtained also in the papers \cite{Gol} by D.Golosov and
A.Ruckenstein.  For the treatment of this problem in a Fermi liquid
the Matthiessen-type rule arguments and simple relaxation-time
approximation for the collision integral have been used \cite{Mat}. 
More recently, the derivation of transverse spin dynamics in a
spin-polarized Fermi liquid from the Landau-Silin kinetic equation
with general form of a two-particle collision integral has been
performed \cite{Min}.  The existance of zero-temperature damping of
transverse spin waves has been established.  At low temperatures and
polarizations $\gamma H$ it proves to be proportional to the rate of
collisions between quasiparticles
\begin{equation}
\frac{1}{\tau}~ \propto~ ((\gamma H)^{2}¥+(2\pi T)^{2}¥).
\label{1}
\end{equation}

Experimentally the saturation of the transverse spin wave
diffusion constant at temperatures about several millikelvin has been
registered by the spin-echo technique (see, for instance,
\cite{Akim}).  On the other hand, the spin wave experiments
demonstrate the behaviour characterized rather by the absence of
transverse spin wave damping in the same temperature region \cite{Ver}. 
The latter seem to be a confirmation of the point of view of I.Fomin
\cite{Fom} who has argued for the dissipationless form of transversal
spin wave spectrum derived from the correction to the system energy
due to the gauge transformation into the coordinate system where the
magnetization vector is constant. 
The calculation of the generalized
susceptibility coefficient in the expression for the spin current
found in \cite{Fom} has not been performed, just the reference on such
calculation \cite{Maki} in superfluid $^{3}¥He$ had been given.  Indeed,
one can calculate susceptibility using a similar procedure.  However,
in the case of polarized Fermi liquid one must use the Green functions
with the finite imaginary self-energy parts due to collisions between
quasiparticles as it was done in \cite{Me,Mu}, that inevitably leeds to
the spin waves attenuation.

The derivation of linear on transversal deviations spin wave dynamics has been
undertaken in the paper \cite{Min} at finite polarization.  However,
all the Fermi
liquid chararteristic parameters  have been taken as constants
independent of polarization.  The derivation partly free of this assumption
(taking in consideration the polarization dependence of Fermi
velocity and the density of states)
is proposed in the present
article (Section II).  It results not only in  equations for the
time-space variations of spin and spin current densities with more
general expressions for all the coefficients but reveals the possibility to
find the distinctions for the spin dynamics in spin polarized
paramagnetic Fermi liquid and in ferromagnetic Fermi liquid with
spontaneous magnetization.  
In both systems transverse spin wave
frequency is found to be proportional to $k^{2}¥$ with complex
constant of prortionality such that the damping has a finite value at
$T=0$.

It is well known that the phenomenological Landau Fermi liquid theory
has well established foundations based on microscopic theory.  Namely,
the transport equation for the vibrations of a Fermi liquid was
derived from an integral equation for vertex function and general
relation between the amplitude of forward scattering and the Fermi liquid 
interaction parameters was found \cite{Landau}.  On the other hand
there were known several publications where the kinetic equation and
field theoretical methods based on Landau Fermi liquid theory have
been applied to the treatment of itinerant isotropic ferromagnet
\cite{AbrDz, Kondr, DzKondr}.  In particular the derivation of
dissipationless (up to the terms of order $\sim k^{4}¥$ ) spin waves
spectrum has been announced \cite{Kondr}.

In the present article in frame of microscopic theory we reconsider
the problem of transverse spin waves in spin-polarized Fermi liquid
(Section III) and in itinerant ferromagnet taking into account the
divergency of static susceptibility (Section IV).  It is shown that in
the both cases the microscopic derivation leads to the same spin wave
spectrum.  Along with the dissipationless part it contains also the
finite zero-temperature damping.  The polarization dependence both
dissipative and reactive part of diffusion constant corresponds to
dependences found by means of kinetic equation with two-particle
collision integral (Section II).

\section{Spin wave dispersion}

The quasiparticle distribution function as well as quasiparticle
energy are given by $2\times2$ matrix in spin space,
\begin{equation}
\hat n_{\bf k}¥({\bf r},t)=n_{\bf k}¥({\bf r},t)\hat I +
\mbox{\boldmath $\sigma $}_{\bf k}¥({\bf r},t) \mbox{\boldmath
$\hat \sigma $},
\label{e2}
\end{equation}
\begin{equation}
\hat \varepsilon_{\bf k}¥({\bf r},t)=
\varepsilon_{\bf k}¥({\bf r},t)\hat I +
{\bf h}_{\bf k}¥({\bf r},t) \mbox{\boldmath$\hat \sigma$}.
\label{e3}
\end{equation}
Here $\mbox{\boldmath$\hat\sigma$}$= $({\hat\sigma}_{x}¥,
{\hat\sigma}_{y}¥,{\hat\sigma}_{z}¥)$ are Pauli matrices. 
As long
as we consider the small deviations of the magnetization direction from
its equilibrium direction
the equation for the scalar part of the
distribution function $n_{\bf k}¥({\bf r},t)$ 
decouples from the equation for the vector part of distribution function
$\mbox{\boldmath $\sigma $}_{\bf k}¥({\bf r},t)$ and we may put
$n_{\bf k}¥$ equal to its equilibrium value, namely, usual Fermi
function. 
Hence, the equation for the $\mbox{\boldmath $\sigma $}_{\bf k}¥({\bf
r},t)$ has the form
\begin{eqnarray}
&&\frac{\partial \mbox{\boldmath$\sigma$}_{\bf k}¥}{\partial t} +
\frac{\partial\varepsilon_{\bf k}¥}
{\partial k_{i}¥}
\frac{\partial \mbox{\boldmath$\sigma$}_{\bf k}¥}{\partial x_{i}¥}-
\frac{\partial {\bf h}_{\bf k}¥}
{\partial x_{i}¥} \frac{\partial n_{\bf k}¥}{\partial k_{i}¥}-
2({\bf h}_{\bf k}¥\times \mbox{\boldmath$\sigma$}_{\bf k}¥) \nonumber \\
&=&\left (\frac{\partial \mbox{\boldmath$\sigma$}_{\bf k}¥}{\partial t}
\right )_{coll}¥.
\label{e4}
\end{eqnarray}
We divide all matrices in equilibrium and nonequilibrium parts,
\begin{equation}
\hat n_{\bf k}¥=\hat n_{\bf k}¥^{0}¥+ \delta \hat n_{\bf k}¥ ,
\label{e5}
\end{equation}
\begin{equation}
\hat \varepsilon_{\bf k}¥= \hat \varepsilon_{\bf k}¥^{0}¥ +
\delta \hat \varepsilon_{\bf k}¥,
\label{e6}
\end{equation}
where
\begin{equation}
\hat n_{\bf k}¥^{0}¥={\bar n}_{0}¥(\varepsilon_{\bf k}¥)\hat I+ 
\frac{1}{2}\Delta n_{0}¥(\varepsilon_{\bf k}¥)(\hat {\bf
m}\mbox{\boldmath$\hat\sigma$})
\label{e7}
\end{equation}
is the equilibrium distribution function of polarized Fermi liquid
and
\begin{equation}
\hat \varepsilon_{\bf k}¥^{0}¥ =\varepsilon_{\bf k}¥\hat I
-\frac{1}{2}\gamma({\bf B}\mbox{\boldmath$\hat\sigma$}) 
\label{e8}
\end{equation}
is the equilibrium quasiparticle energy.
Here, the functions
\begin{equation}
{\bar n}_{0}¥(\varepsilon_{\bf k}¥)=\frac{1}{2}(n_{0}¥^{+}¥+n_{0}¥^{-}¥)
\label{e9}
\end{equation}
and
\begin{equation}
\Delta n_{0}¥(\varepsilon_{\bf k}¥)=n_{0}¥^{+}¥-n_{0}¥^{-}¥
\label{e10}
\end{equation}
are determined through two Fermi distribution functions
\begin{equation}
n_{0}¥^{\pm}¥(\varepsilon_{\bf k}¥)=
n_{0}¥(\varepsilon_{\bf k}¥\mp \frac{\gamma H}{2})=
\frac{1}{\exp\left(\frac {\varepsilon_{\bf k}¥\mp \frac{\gamma
H}{2} -\mu }{T}\right ) +1}
\label{e11}
\end{equation}    
shifted on the value of polarization $\gamma H/2$, $\gamma$ is the gyromagnetic ratio,
Planck constant $\hbar=1$
throughout the paper, the polarization direction is determined by the
unit vector ${\bf m}= {\bf H}/H$.

We have introduced two magnetic fields  ${\bf H}$ and ${\bf B}$ and 
shall assume that they are parallel each other.  The field ${\bf H}$
determining the shift of the quasiparticle distribution function
corresponds in paramagnetic Fermi liquid to the magnetization created
by the external magnetic field ${\bf H}_{0}¥$ and by the pumping
\cite{Rod}.   The pumped part in view of very
long time of longitudinal relaxation should be considered as
equilibrium part of magnetization.  
In a ferromagnetic Fermi liquid ${\bf H}$ is spontaneous
magnetic field existing even in absence of an external field and
pumping.  The field ${\bf B}$ determines the shift in energy of
quasiparticles consisting of an external magnetic field ${\bf H}_{0}¥$
and the Fermi-liquid molecular field.  To define ${\bf B}$ we must
consider the equilibrium distribution matrix (\ref{e7}) and
equilibrium energy matrix (\ref{e8}) as deviations from the
corresponding matrices for nonpolarized Fermi liquid,
\begin{equation}
\hat n_{\bf k}¥^{0}¥=n_{0}¥(\varepsilon_{\bf k}¥)\hat I+ \delta
\hat n_{\bf k}¥^{0}¥,
\label{e12}
\end{equation}
\begin{eqnarray}
&&\hat \varepsilon_{\bf k}¥^{0}¥=\varepsilon_{{\bf k}}¥\hat I
-\frac{1}{2}\gamma({\bf B}\mbox{\boldmath$\hat\sigma$})\nonumber \\
&=&
\varepsilon_{{\bf k}}¥\hat I -\frac{1}{2}\gamma({\bf
H}_{0}¥\mbox{\boldmath$\hat\sigma$})+ \frac{1}{2}Sp'\int d\tau'
f_{{\bf k}{\bf k'}}¥^{\sigma \sigma'}¥ \delta \hat n_{\bf k'}¥^{0}¥,
\label{e13}
\end{eqnarray}
where $d \tau=2d {\bf k}/(2\pi)^{3}¥$ and $f_{{\bf k}{\bf k'}}¥^{\sigma \sigma'}¥$ 
is the Fermi-liquid interaction matrix.

As it was discussed in \cite{Min} for the finite polarization and  the
general form of $f_{{\bf k}{\bf k'}}¥^{\sigma \sigma'}¥$ the vector
$\bf B$ proves to be energy dependent.  This in its turn leads to impossibility
of the spin dynamics description in terms of two closed equtions for
spin and spin current densities.
To circumvent these difficulties as in the paper \cite{Min} we assume the
independence functions $f_{{\bf k}{\bf k'}}$ of energy and 
take them in the simplified form
\begin{equation}
f_{{\bf k}{\bf k'}}¥^{\sigma \sigma'}¥=
f_{{\bf k}{\bf k'}}¥^{s}¥\hat I\hat I'+
[f_{0}¥^{a}¥+f_{1}¥^{a}¥(\hat{\bf k}\hat{\bf k'})] \mbox{\boldmath$\hat \sigma
$}\mbox{\boldmath$\hat \sigma' $}.
\label{e14}
\end{equation}    
Now, from (\ref{e13}), (\ref{e14}) we obtain an equation for ${\bf B}$
determination
\begin{equation}
\gamma{\bf B}=\gamma{\bf H}_{0}¥- \hat{\bf m} f_{0}¥^{a}¥ \int d\tau
\Delta n_{0}¥.
\label{e15}
\end{equation}
For small polarizations and taking for simplicity $T=0$ one can rewrite (\ref{e15})
as
\begin{equation}
{\bf B}={\bf H}_{0}¥-{\bf H}F_{0}¥^{a}¥\left(1-
\frac{1}{6}\left (\frac{\gamma H}{4\mu}\right)^{2}¥\right ) .
\label{e16}
\end{equation}
Here $F_{0}¥^{a}¥=N_{0}¥f_{0}¥^{a}¥$,  
$N_{0}¥=m^{*}¥k_{F}¥/\pi^{2}¥$ is the density of states at zero
polarization.  In the absence of a pumped magnetization the field ${\bf
B}={\bf H}$ and (\ref{e16}) is just the self-consistency equation for
the field ${\bf H}$ determination as the function of an external field
${\bf H}_{0}¥$ giving in the lowest order
\begin{equation}
{\bf H}= \frac{{\bf H_{0}¥}}{1+F_{0}¥^{a}¥}.
\label{e17}
\end{equation}
As a particular case 
one can consider also a ferromagnetic state
realized at $F_{0}¥^{a}¥=-1-\delta $ when the solution of the equation
(\ref{e16})
\begin{equation}
\frac{1}{6}\left (\frac{\gamma H}{4\mu}\right)^{2}¥=\frac{1+F_{0}¥^{a}¥}
{F_{0}¥^{a}¥}
\label{e18}
\end{equation}
exists even in the absence of an external field.

When the part of magnetization is
created by pumping, ${\bf H}$ presents an independent value and the
total energy shift $\gamma ({\bf B}\mbox{\boldmath$\hat \sigma $})/2$
is determined by means of two fields: external ${\bf H}_{0}¥$ and
"effective" ${\bf H}$. 

We discuss the only perpendicular deviations from the initial
equilibrium state,
\begin{equation}
\delta \hat n_{\bf k}¥=
\delta\mbox{\boldmath $\sigma $}_{\bf k}¥({\bf r},t) \mbox{\boldmath
$\hat \sigma $},  ~~~(\hat {\bf m}\delta\mbox{\boldmath $\sigma $}_{\bf k}¥)=0.
\label{e19}
\end{equation}    
Then the energy deviation matrix  has the form
\begin{equation}
\delta\hat \varepsilon_{\bf k}¥ =\delta{\bf h}_{\bf k}¥
\mbox{\boldmath$\hat \sigma$}, ~~~
\delta{\bf h}_{\bf k}¥=
\int d\tau'f_{{\bf k}{\bf k'}}¥^{a}¥ \delta\mbox{\boldmath
$\sigma $}_{\bf k'}¥
\label{e20}
\end{equation}
and the kinetic equation (\ref{e4}) can be rewritten as
\begin{eqnarray}
&&\frac{\partial \delta \mbox{\boldmath$\sigma$}_{\bf k}¥}{\partial t} +
\frac{\partial\varepsilon_{\bf k}¥^{0}¥} {\partial
k_{i}¥}
\frac{\partial \delta
\mbox{\boldmath$\sigma$}_{\bf k}¥}{\partial x_{i}¥}-
\frac{\partial \bar n_{0}¥} {\partial k_{i}¥}\frac{\partial \delta{\bf
h}_{\bf k}¥}{\partial x_{i}¥}\nonumber \\
&-&2\left [\left(-\frac{1}{2}\gamma{\bf B}+\delta{\bf h}_{\bf
k}¥\right)\times \left(\frac{1}{2} {\hat {\bf m}}\Delta n_{0}\hat +\delta
\mbox{\boldmath$\sigma$}_{\bf k}¥ \right)\right ]\nonumber \\
&=&\left
(\frac{\partial \mbox{\boldmath$\sigma$}_{\bf k}¥}{\partial t} \right
)_{coll}¥.
\label{e21}
\end{eqnarray}

To derive the closed system of equations for the spin density ${\bf
M}$ and the spin current density ${\bf J}_{i}¥$ in the case of finite
polarization we make
an assumption which is plausible for weakly polarized Fermi liquid
that the energy dependence of $\delta\mbox{\boldmath $\sigma $}_{\bf
k}¥({\bf r},t)$ is factorized from the space and direction of ${\bf
\hat k}$ dependences:
\begin{equation}
\delta\mbox{\boldmath $\sigma $}_{\bf k}¥({\bf r},t)=
{\bf A}({\bf r},t)\alpha (\varepsilon)+{\bf B}_{i}¥({\bf
r},t)\hat k_{i}¥\beta (\varepsilon).
\label{e22}
\end{equation}
In terms of these functions one can write the spin density
\begin{equation}
{\bf M}({\bf r},t)=
\frac{1}{2}\int d \tau \delta \mbox{\boldmath$\sigma$}_{\bf k}¥
=\frac{1}{2}{\bf A}({\bf r},t)\int d \tau~\alpha (\varepsilon),
\label{e23}
\end{equation}
and spin current density
\begin{eqnarray}
{\bf J}_{i}¥({\bf r},t)=\frac{1}{2}\int d \tau \left [v_{i}¥\delta
\mbox{\boldmath$\sigma$}_{\bf k}¥-\frac{\partial n_{0}¥}{\partial
k_{i}¥} \delta{\bf h}_{\bf k}¥\right ] = \frac{1}{2}\psi\int d
\tau v_{i}¥\delta \mbox{\boldmath$\sigma$}_{\bf k}¥&&\nonumber \\
= \frac{1}{6}{\bf B}_{i}¥({\bf r},t)
\int d \tau v_{i}(\varepsilon)¥ [\beta
(\varepsilon)-\frac{f_{1}¥^{a}¥}{3}\frac{\partial \bar n_{0}¥}
{\partial \varepsilon}\int d \tau \beta (\varepsilon)],
\label{e24}~~~~&&
\end{eqnarray}
where $v_{i}(\varepsilon)¥=\frac{\partial \varepsilon_{\bf k}¥}{
\partial k_{i}¥}$ and
\begin{equation}
\psi=\frac{\int d \tau v(\varepsilon)¥ [\beta
(\varepsilon)-\frac{f_{1}¥^{a}¥}{3}\frac{\partial \bar n_{0}¥}
{\partial \varepsilon}\int d \tau \beta (\varepsilon)]}{\int d\tau
v(\varepsilon)\beta(\varepsilon)}
\label{e25}
\end{equation}

Making the integrations of kinetic equation (\ref{e21})
$\int d \tau/2 $ and $\int d \tau v_{i}¥/2$ we obtain
\begin{equation}
\frac{\partial {\bf M}}{\partial t} +\frac{\partial 
{\bf J}_{i}¥}{\partial x_{i}¥} -{\bf M}\times \gamma {\bf H}_{0}¥=0,
\label{e26}
\end{equation}
\begin{eqnarray}
\frac{\partial {\bf J}_{i}¥}{\partial t} &+&
\frac{w^{2}¥}{3}
\frac{\partial{\bf M}}{\partial x_{i}¥} - {\bf J}_{i}¥\times\gamma{\bf
H}_{0}¥ + {\bf J}_{i}¥
\times{\bf C}\nonumber \\
&=&\frac{\psi}{2}\int d \tau v_{i}¥ \left (\frac{\partial
\mbox{\boldmath$\sigma$}_{\bf k}¥}{\partial t} \right )_{coll}¥.
\label{e27}
\end{eqnarray}
Here
\begin{equation}
w^{2}¥=\psi\left[\frac{\int d\tau v^{2}¥(\varepsilon)
\alpha(\varepsilon)}{\int d\tau \alpha(\varepsilon)}- f_{0}¥^{a}¥\int d\tau
v^{2}¥(\varepsilon)\frac{\partial\bar n_{0}¥}{\partial \varepsilon}\right]
\label{e28}
\end{equation}
and
\begin{equation}
{\bf C}=\hat{\bf m}\left[f_{0}¥^{a}¥\int d\tau \Delta n_{0}¥(\varepsilon)
-\frac{f_{1}¥^{a}¥\int d\tau \beta(\varepsilon)\int d\tau
v(\varepsilon) \Delta n_{0}¥(\varepsilon)}{3\int d\tau \beta(
v(\varepsilon)\varepsilon)}\right] .
\label{e29}
\end{equation}

The equations (\ref{e26}), (\ref{e27}) have the same form as in the case of
vanishingly small polarization \cite{Leg,Min}.  The equation
(\ref{e27}) is transformed to the analogous equation for vanishingly
small polarization
\begin{eqnarray}
\frac{\partial {\bf J}_{i}¥}{\partial t} &+&
\frac{1}{3}v_{F}¥^{2}¥(1+F_{0}¥^{a}¥)(1+\frac{F_{1}¥^{a}¥}{3})
\frac{\partial{\bf M}}{\partial x_{i}¥}\nonumber \\ 
&-& {\bf J}_{i}¥\times\gamma{\bf
H}_{0}¥ +\frac{4}{N_{0}¥}(F_{0}¥^{a}¥-\frac{F_{1}¥^{a}¥}{3})({\bf J}_{i}¥
\times{\bf M}^{\parallel}¥)\nonumber \\
&=&\frac{1}{2}(1+\frac{F_{1}¥^{a}¥}{3})\int d \tau v_{Fi}¥
\left (\frac{\partial
\mbox{\boldmath$\sigma$}_{\bf k}¥}{\partial t} \right )_{coll}¥,
\label{e30}
\end{eqnarray}
\begin{equation}
{\bf M}^{\parallel}¥=
\frac{\gamma N_{0}¥}{4}{\bf H},~~~F_{i}¥^{a}¥=N_{0}¥f_{i}¥^{a}¥
\label{e31}
\end{equation}
if
we put
\begin{equation}
\alpha(\varepsilon)~\propto~\beta (\varepsilon)~\propto~ 
\Delta n_{0}¥(\varepsilon).
\label{e32}
\end{equation}
Thus, one can work with eqn (\ref{e27}) 
taking more specific definitions for
\begin{equation}
\psi=\frac{\int d \tau v(\varepsilon)¥
[\Delta n_{0}¥(\varepsilon)-\frac{f_{1}¥^{a}¥}{3}\frac{\partial \bar n_{0}¥} {\partial
\varepsilon}\int d \tau \Delta n_{0}¥(\varepsilon)]}{\int d\tau
v(\varepsilon)\Delta n_{0}¥(\varepsilon)},
\label{e33}
\end{equation}
\begin{equation}
w^{2}¥=\psi\left[\frac{\int d\tau v^{2}¥(\varepsilon)
\Delta n_{0}¥(\varepsilon)}{\int d\tau \Delta n_{0}¥(\varepsilon)}- f_{0}¥^{a}¥\int d\tau
v^{2}¥(\varepsilon)\frac{\partial\bar n_{0}¥}{\partial \varepsilon}\right],
\label{e34}
\end{equation}
and
\begin{equation}
{\bf C}=\frac{\hat{\bf m}}{N_{0}¥}(F_{0}¥^{a}¥ -\frac{F_{1}¥^{a}¥}{3})\int
d\tau \Delta n_{0}¥(\varepsilon) .
\label{e35}
\end{equation}
At last, using the calculations of the paper \cite{Min} for the collision
integral in weakly polarized liquid we come to the equation for the spin
current density
\begin{equation}
\frac{\partial {\bf J}_{i}¥}{\partial t} +
\frac{w^{2}¥}{3}
\frac{\partial{\bf M}}{\partial x_{i}¥} - {\bf J}_{i}¥\times\gamma{\bf
H}_{0}¥ + {\bf J}_{i}¥
\times{\bf C}=-\frac{{\bf J}_{i}¥}{\tau}
\label{e36}
\end{equation}
where the current relaxation time is
\begin{equation}
\frac{1}{\tau}=\frac{m^{*}¥^{3}¥}{6(2\pi)^{5}¥} (2\overline {W_{1}¥}+\overline{W_{2}¥})
\left[(2\pi T)^{2}¥+(\gamma H)^{2}¥\right].
\label{e37}
\end{equation}
The  dispersion law of the transversal spin waves following from
equations (\ref{e26}), (\ref{e36}) is (see for instance \cite{Can})
\begin{equation}
\omega=\omega_{L}¥+ (D^{\prime\prime}¥-iD^{\prime}¥)k^{2}¥,
\label{e38}
\end{equation}
where $\omega_{L}¥=\gamma H_{0}¥$ is the Larmor frequency,
\begin{equation} 
D^{\prime}¥=\frac{w^{2}¥\tau} {3(1+(C \tau)^{2}¥
)}\cong\frac{w^{2}¥}{3{C}^{2}¥ \tau}
\label{e39}
\end{equation}
is the dissipative part of diffusion coefficient and 
\begin{equation} 
D^{\prime\prime}¥= C\tau D^{\prime}¥\cong\frac{w^{2}¥}{3 C}
\label{e40}
\end{equation}
is its reactive part.  Here the second approximative values of $D^{\prime}¥$
and $D^{\prime\prime}¥$ correspond to the limit $C\tau\gg1$.

For
a weakly polarized fluid $C=(F_{0}¥^{a}¥-F_{1}¥^{a}¥/3)\gamma H$ and 
$\psi=1+F_{1}¥^{a}¥/3$.
The expression for $w^{2}¥$ depends of state of liquid.  One can find
it analitically in the case of weak polarization.
In a paramagnetic Fermi liquid it is
\begin{equation} 
w^{2}¥=v_{F}¥^{2}¥(1+F_{0}¥^{a}¥)(1+\frac{F_{1}¥^{a}¥}{3})
\label{e40a}
\end{equation}
where $v_{F}$ is the Fermi velocity in nonpolarized liquid. In a
ferromagnetic Fermi liquid (if an external field is smaller than
spontaneous) we find from eqn (\ref{e34}) with help of eqn (\ref{e15})
\begin{equation} 
w^{2}¥=-v_{F}¥^{2}¥(1+\frac{F_{1}¥^{a}¥}{3})
\left(\frac{\gamma H}{4\mu}\right)^{2}¥
\label{e40b}
\end{equation}

Thus, the reactive part of diffusion coefficient in paramagnetic state
at $T=0$ proves to be inversely proportional to magnetization
\begin{equation} 
D^{\prime\prime}¥=\frac{v_{F}¥^{2}¥(1+F_{0}¥^{a}¥)(1+F_{1}¥^{a}¥/3)}
{3 (F_{0}¥^{a}¥-F_{1}¥^{a}¥/3)\gamma H}
\label{e40c}
\end{equation}
 whereas in ferromagnetic state it is directly proportional to
magnetization
\begin{equation} 
D^{\prime\prime}¥=-\frac{v_{F}¥^{2}¥(1+F_{1}¥^{a}¥/3)\gamma H} {3
(F_{0}¥^{a}¥-F_{1}¥^{a}¥/3)(4\mu )^{2}¥}.
\label{e40d}
\end{equation}
The
latter is in correspondence with known result obtained in frame of
Stoner-Hubbard model \cite{Mor}.  

The dissipative part of diffusion
coefficient given by eqn (\ref{e39}) at $T=0$ in paramagnetic state is polarization
independent, whereas in ferromagnetic state it is proportional to
square of magnetization.

So, the transverse spin waves frequency in polarized paramagnetic
Fermi liquid as well in a Fermi liquid with spontaneous magnetization is
found to be proportional to $k^{2}¥$ with complex diffusion
coefficient such that the damping at $C\tau\gg 1$ has a finite value
proportional to the scattering rate of quasiparticles at $T=0$.  As it
was pointed out in \cite{Min} the latter is in formal analogy with
ultrasound attenuation in collisionless regime.  It is worth noting,
however, that in neglect of processes of longitudinal relaxation the 
parameter $\gamma H \tau$ has no relation to the local equilibrium
establishment.

The results (\ref{e38})-(\ref{e40d}) are valid both in hydrodynamic
$Dk^{2}¥\tau \ll 1$ and in collisionless regime $ Dk^{2}¥\tau\gg 1$ so
long
\begin{equation} 
Dk^{2}¥\ll \gamma H
\label{e40e}
\end{equation}
that is the condition
of two moment approximation (\ref{e22}) for the solution of the
kinetic equation \cite{Leg}.  This behavior of polarized Fermi
liquid contrasts with the behavior of Heisenberg ferromagnet in
hydrodynamic regime \cite{Hal} where the transverse spin wave
attenuation appears in terms proportional $k^{4}¥$.

\section{Microscopic derivation of spin wave spectrum in polarized
Fermi liquid}

The Landau-type derivation of transverse spin
dynamics in a weakly spin-polarized Fermi-liquid from microscopic
theory has been performed in the paper \cite{Mu}.  Here we make a
similar derivation with the purpose to stress the conditions it needs
to be valid, to compare the answer with that obtained from kinetic
equation at nonzero temperatures, and to juxtapose this with the
derivation for ferromagnetic Fermi-liquid \cite{DzKondr} which we also
reproduce afterwards.

As in the original paper by Landau \cite{Landau} we consider a system of
fermions at $T=0$, with arbitrary short range interaction forces.  The
presence of polarization means that subsystems of spin-up and spin-down
particles have different chemical potentials
$\mu_{\pm}¥=\mu\pm \gamma B/2$ and the distribution functions with
different Fermi momenta $p_{\pm}¥=p_{0}¥\pm \gamma H/2v_{F}¥$.  The
polarization in general is nonequilibrium and, as in previous section, we
shall distiguish the fields ${\bf H}$ and ${\bf B}$.  Here, we shall
not take in mind the polarization dependence of the Fermi velocity and
density of states and obtain results relating to weakly polarized
paramagnetic Fermi liquid.  The ferromagnetic case shall be
discussed in the next section.
So, the Fermi velocity is $ v_{F}¥=\frac{\partial \varepsilon(p)}{
\partial p}|_{p=p_{0}¥}¥ $ and $p_{0}¥=\frac{p_{+}¥+p_{-}¥}{2}$.  The
Green functions near $|{\bf p}|=p_{\pm}¥$ and $\varepsilon({\bf
p})=\mu_{\pm}¥$ have the form
\begin{equation}
G_{\pm}¥({\bf p}, \varepsilon)=\frac{a}{\varepsilon
-\varepsilon({\bf p})+\mu_{\pm}¥
+ibv_{F}¥^{2}¥(p-p_{\pm}¥)|p-p_{\pm}¥|}.
\label{e41}
\end{equation}

We use a weak polarization $v_{F}¥(p_{+}¥-p_{-}¥)\ll\varepsilon_{F}¥$
and also assume that both the Fermi distributions are characterized by the same
Landau Fermi liquid parameters.  We introduce here the general form
of imaginary part of self-energy \cite{LifPit} which is quadratic function
of the difference $(p-p_{\pm}¥)$ and changes its sign at $p=p_{\pm}¥$
correspondingly.  The assumption of small polarization means in
particular that $G_{+}¥$ is given by the expression (\ref{e41}) not
only near 
$|{\bf p}|=p_{+}¥$ and
$\varepsilon({\bf p})=\mu_{+}¥$ but in the whole intervals $p_{-}¥<p<p_{+}¥$ and
$\mu_{-}¥<\varepsilon({\bf p})<\mu_{+}¥$ and also near  $|{\bf
p}|=p_{-}¥$ and $\varepsilon({\bf p})=\mu_{-}¥$.  The same is true for $G_{-}¥$.

Following Landau, let us write the equation for the vortex function 
describing
scattering of two particles with opposite spin directions and a small
transfer of 4-momentum $K=({\bf k},\omega)$
\begin{eqnarray}
&&\Gamma(P_{1}¥,P_{2}¥,K)=\Gamma_{1}¥(P_{1}¥,P_{2}¥)
-\frac{i}{(2\pi)^{4}¥} \int
\Gamma_{1}¥(P_{1}¥,Q)\nonumber \\
&\times& G_{+}¥(Q)G_{-}¥(Q+K) \Gamma(Q,P_{2}¥,K)d^{4}¥Q
\label{e42}
\end{eqnarray}
If $K$ is small and polarization is also small, the poles of two Green
functions are close to each other.  Let us assume that all other
quantities in the integrand are slowly varying with respect to $Q$:
their energy and momentum scales of variation are larger than $\max
\{\gamma H, \omega\}$ and $\max\{\gamma H/v_{F}¥,k\}$ correspondingly.  Then
one can perform the integration in (\ref{e42}) at fixed values of
$q=p_{0}¥,~ \varepsilon=0$ in the arguments of $\Gamma$ and $\Gamma_{1}¥$
functions.  In other words, one can substitute in (\ref{e42})
\begin{eqnarray}
&&G_{+}¥(Q)G_{-}¥(Q+K)=G_{+}¥({\bf q},\varepsilon)
G_{-}¥({\bf q}+{\bf k},\varepsilon+\omega) \nonumber \\
&=&\frac{2\pi i a^{2}¥}{v_{F}¥}\delta(\varepsilon)\delta(|{\bf q}|-p_{0}¥)
\nonumber \\
&\times&\frac{\gamma H+{\bf k}{\bf v}_{F}¥}{\omega-\omega_{L}¥
+\gamma H F_{0}¥^{a}¥+ib(\gamma H)^{2}¥/2- {\bf k}{\bf v}_{F}¥+ ib{\gamma H\bf
k}{\bf v}_{F}¥}\nonumber \\
&+&
\Phi_{\mbox{reg}}¥.
\label{e43}
\end{eqnarray}
For eliminating $\Gamma_{1}¥$ from (\ref{e2}) we shall rewrite this
equation in the operator form
\begin{equation}
\Gamma=\Gamma_{1}¥-i\Gamma_{1}¥(i\Phi+\Phi_{\mbox{reg}}¥)\Gamma,
\label{e44}    
\end{equation}
where product is interpreted as integral, and $i\Phi$ denotes the
first term from right-hand side eq.  (\ref{e43}).  In equation (\ref{e44}), we
transpose the term involving $\Phi_{\mbox{reg}}¥$ to the left-hand
side, and then apply the operator
$(1+i\Gamma_{1}¥\Phi_{\mbox{reg}}¥)^{-1}¥$, obtaining
\begin{equation}
\Gamma=\Gamma^{\omega}¥+\Gamma^{\omega}¥\Phi\Gamma,
\label{e45}
\end{equation}
where
\begin{equation}
\Gamma^{\omega}¥
=(1+i\Gamma_{1}¥\Phi_{\mbox{reg}}¥)^{-1}¥\Gamma_{1}¥.
\label{e46}
\end{equation}
As it is known \cite{Landau}, $\Gamma^{\omega}¥( H=0)$ is directly
related to the function determining the Fermi liquid interaction,
\begin{equation}
\Gamma^{\omega}¥(H=0)=\Gamma((|{\bf k}|/\omega)\to 0, H=0)=
\frac{F_{{\bf n}{\bf n'}}}{a^{2}¥N_{0}¥}.
\label{e47}
\end{equation}
At finite $H$ the $\Gamma^{\omega}¥$ function can be expanded over
the polarization as
\begin{equation}
{a^{2}¥N_{0}¥}\Gamma^{\omega}¥={F_{{\bf n}{\bf n'}}} +
ib\gamma H C_{{\bf n}{\bf n'}} +O(H^{2}¥).
\label{e48}
\end{equation}
From eqns (\ref{e45}) and (\ref{e48}), we come, according to a well known
procedure \cite{Landau}, to kinetic equation
\begin{eqnarray}
\left(\omega -\omega_{L}¥+\gamma H F_{0}¥^{a}¥ +\frac{ib(\gamma
H)^{2}¥}{2} -{\bf k}{\bf n}v_{F}¥\right. &&~~~~\nonumber \\
 + ib{\bf k}{\bf n} v_{F}¥\gamma H 
\left .\right)
\nu({\bf n})&& \nonumber \\
= \left (\gamma H+{\bf k}{\bf n}v_{F}¥\right) {\displaystyle \int}
\frac{d{\bf n'}¥}{4\pi} \left(F_{{\bf n}{\bf n'}}+ ib\gamma H C_{{\bf
n}{\bf n'}}\right)\nu({\bf n'}).  \label{e49}&&
\end{eqnarray}
We limit ourself to the first two harmonics in the Landau interaction
function $F_{{\bf n}{\bf n'}}=F_{0}¥^{a}¥+({\bf n}{\bf
n'})F_{1}¥^{a}¥$ and $C_{{\bf n}{\bf n'}}=C_{0}¥+({\bf n}{\bf
n'})C_{1}¥$.  To obtain the spectrum of the spin waves (see below)
obeying the Larmor theorem: the system of spins in a homogeneous
magnetic field executes the precessional motion with the Larmor
frequency $\omega_{L}¥=\gamma H_{0}¥$, the coefficient $C_{0}¥$ has to
be chosen \cite{Ward} equal to 1/2 .

Introducing the expansion of the distribution
function $\nu({\bf n})¥$ over spherical harmonics of direction ${\bf
n}={\bf v}_{F}¥/v_{F}¥$, one can find from (\ref{e49}) that the ratio of
amplitudes of the successive harmonics with $l\ge 1$ is of the order
of $kv_{F}¥/\gamma H$.  Hence if this ratio is assumed to be a small
parameter one can work with distribution function taken in the form
\cite{Leg} $\nu({\bf n})=\nu_{0}¥+({\bf n}\hat{\bf k})\nu_{1}¥$.  The
functions $\nu_{0}¥$ and $\nu_{1}¥$ obey the following system of
linear equations:
\begin{equation}
( \omega-\omega_{L}¥ )\nu_{0}¥-\frac{kv_{F}¥}{3}
\left (1+\frac{F_{1}¥^{a}¥}{3}
-ib(1-\frac{C_{1}¥}{3})\gamma H \right ) \nu_{1}¥=0,
\label{e50}
\end{equation}
\begin{eqnarray}
-kv_{F}¥ \left ( 1+F_{0}¥^{a}¥-\frac{ib\gamma H}{2} \right )\nu_{0}¥ &&
\nonumber\\
+\left(\omega-\omega_{L}¥+(F_{0}¥^{a}¥-\frac{F_{1}¥^{a}¥}{3})\gamma H  +
ib(\frac{1}{2}-\frac{C_{1}¥}{3})(\gamma H)^{2}¥\right )\nu_{1}¥&&\nonumber\\
=0.~~~~~~~~~~~~~~~~~~~~~~~~~~~~~~~~~~~~~~~&&
\label{e51}
\end{eqnarray}
Vanishing of the determinant of this system gives the spin waves dispersion
law.  At long enough wave-lengths when the dispersive part of
$\omega(k)$ dependence is much less than $\omega_{L}¥$ and neglecting
the terms $\sim (b\gamma H)^{2}¥$, we have
\begin{equation}
\omega=\omega_{L}¥+ (D^{\prime\prime}¥-iD^{\prime}¥)k^{2}¥,
\label{e52}
\end{equation}
where 
\begin{equation} 
D^{\prime\prime}¥=\frac{v_{F}¥^{2}¥(1+F_{0}¥^{a}¥)(1+F_{1}¥^{a}¥/3)}
{3(F_{0}¥^{a}¥-F_{1}¥^{a}¥/3) \gamma H}
\label{e53}
\end{equation}
is a reactive part of the diffusion coefficient,
\begin{equation}     
D^{\prime}¥=\frac{bv_{F}¥^{2}¥[(1-C_{1}¥/3)
(1+F_{0}¥^{a}¥)^{2}¥-(1+F_{1}¥^{a}¥/3)^{2}¥/2]}{3(F_{0}¥^{a}¥-
F_{1}¥^{a}¥/3)^{2}¥}
\label{e54}
\end{equation}
is a dissipative part of the diffusion coefficient.  We derived eqns
(\ref {e53}) and (\ref {e54}) in the assumption of 
$(F_{0}¥^{a}¥-F_{1}¥^{a}¥/3) \ne 0$.

The expressions for $D^{\prime\prime}¥$ and $D^{\prime}¥$ have been
obtained first by the same method by A.Meyerovich and K.Musaelyan
\cite{Mu}.  The former is literally coincides with that found in this
paper, the latter has the same parametric dependence but depends in
different way from Fermi liquid parameters.  The reason for this is
not clear at the moment.  These expressions reproduce the
corresponding diffusion constants obtained from phenomenological
Landau-Silin kinetic equation with two-particle collision integral
(see \cite{Min} and previous Section) at arbitrary relation between
polarization and temperature if we put in the latters $T=0$.  In
particular, $D^{\prime}¥$ proves to be polarization independent
whereas $D^{\prime\prime}¥$ is inversely proportional to polarization.

Thus, the general microscopic derivation confirms the statement about
the existance of zero-temperature spin waves attenuation in polarized
Fermi liquid.  The value of the dissipative part of spin diffusion
$D^{\prime}¥$ is determined by the amplitude "b" of the imaginary part of
self-energy.  It originates from collisions between quasiparticles.

\section{Microscopic derivation taking into account the transverse static
susceptibility divergency}

There are several known investigations of an isotropic itinerant
ferromagnetic state as some peculiar type of Fermi liquid.  This
subject was discussed first phenomenologically by A.A.Abrikosov and
I.E.Dzyaloshinskii \cite{AbrDz} and then microsopically by
P.S.Kondratenko \cite{Kondr}.  They did not include in the theory a
finite scattering rate between quasiparticles and as result they
obtained the dissipationless transverse spin wave dispersion law as it
seemed to be in isotropic ferromagnet.  The derivation \cite{AbrDz}
was critisized by C.Herring \cite{Her} who pointed out on the existance
of finite scattering rate.  Later I.E.Dzyaloshinskii and
P.S.Kondratenko \cite{DzKondr} rederived the spin-wave dispersion law
in ferromagnets.  Making use as the starting point the Landau equation
for the vertex function for the scattering of two particles with
opposite spin direction and a small transfer of 4-momentum they have
redefined the product of two Green functions $G_{+}¥G_{-}¥$ in such a
manner that its resonant part was taken equal to zero at $\omega=0$. 
This trick gives a possibility to use the $1/k^{2}¥$ divergency of
transverse static susceptibility, which is an inherent property of
degenerate systems and occurs both in an isotropic ferromagnet and in
spin polarized paramagnetic Fermi-liquid.  The latter of course is
true in the absense of interactions violating total magnetization
conservation.  As in the previous papers \cite{AbrDz,Kondr}, the
authors of \cite{DzKondr} did not introduce a scattering rate in the
momentum space between the Fermi surfaces for the particles with
opposite spins.

Let us see now what kind of modifications appear if we reproduce the
derivation proposed in \cite{DzKondr} with the Green functions (\ref{e41})
taking into account the finite quasiparticle scattering rate in the
whole interval $p_{-}¥<p<p_{+}¥$.  We discuss  an isotropic
ferromagnet at equilibrium ${\bf B}={\bf H}$ first in the absence of
external field.  Following \cite{DzKondr} we write:
\begin{eqnarray}
&G_{+}¥&(Q)G_{-}¥(Q+K)=G_{+}¥({\bf q},\varepsilon)
G_{-}¥({\bf q}+{\bf k},\varepsilon+\omega) \nonumber \\
&=&\frac{2\pi i a^{2}¥}{v_{F}¥}\delta(\varepsilon)\delta(|{\bf q}|-p_{0}¥)
\nonumber \\
&\times&\frac{\omega} {\omega -\gamma H+ ib(\gamma H)^{2}¥/2 - {\bf k}{\bf v}_{F}¥+
ib{\bf k}{\bf v}_{F}¥\gamma H}\nonumber \\
&+&\tilde \Phi_{\mbox{reg}}¥.
\label{e61}
\end{eqnarray}
Now the eqn (\ref{e42}) is written as
\begin{equation}
\Gamma=\Gamma_{1}¥-i\Gamma_{1}¥(i\tilde \Phi+\tilde \Phi_{\mbox{reg}}¥)\Gamma,
\label{e62}
\end{equation}
where $i\tilde \Phi$ denotes the first term from right-hand side eq. 
(\ref{e61}).  The equivalent form of this equation is
\begin{equation}
\Gamma=\Gamma^{\bf k}¥+\Gamma^{\bf k}¥\tilde\Phi\Gamma,
\label{e63}
\end{equation}
where
\begin{equation}
\Gamma^{\bf k}¥=\Gamma\left (\frac{\omega}{|{\bf k}|}\to 0\right )
=(1+i\Gamma_{1}¥\tilde\Phi_{\mbox{reg}}¥)^{-1}¥\Gamma_{1}¥.
\label{e64}
\end{equation}
The isotropic part of $\Gamma^{\bf k}¥$ is proportional to the
 static transverse susceptibility.  Hence it has a singular form
 \cite{DzKondr}
\begin{equation}     
\Gamma^{\bf k}¥\propto~
-\frac{1}
{N_{0}¥ (ck)^{2}¥ }.
\label{e65}
\end{equation}
Here, $c$ is a parameter with the dimensions of length.  One can show by direct
calculation of static transverse susceptibility in ferromagnet
\cite{Mor} that $c$ is polarization independent
\begin{equation}  
c\sim \frac{1}{ p_{0}¥}.
\label{e66}
\end{equation}
At the same time the
similar calculations for polarized paramagnetic Fermi liquid gives the
value of $c$ inversely proportional to polarization
\begin{equation}  
c\sim \frac{v_{F}¥}{\gamma H}
\label{e67}
\end{equation}
such that the divergency (\ref{e65})
disappears in nonpolarized liquid when $\gamma H~\to~0$. 

Substitution of eqn (\ref{e65}) into eqn (\ref{e63}) gives the transverse
spin wave dispersion law
\begin{equation}     
\omega= \gamma H(ck)^{2}¥( 1-\frac{ib\gamma H}{2})
\label{e68}
\end{equation}
which proves to be attenuating.  One can take into consideration a
static external field, by working in the rotating with Larmor
frequency coordinate frame that is equivalent to the substitution
$\omega \to \omega-\omega_{L}¥$ (see also \cite{DzKondr}).  As a
result, we obtain the dispersion law
\begin{equation}     
\omega=\omega_{L}¥+\gamma H(ck)^{2}¥( 1-\frac{ib\gamma H}{2})
\label{e69}
\end{equation}
that has the same form as (\ref{e38}).  Taking into account the relations
(\ref{e66}) and (\ref{e67}) one can make sure that
the polarization dependences of reactive and dissipative part of
diffusion constant in ferromagnetic Fermi liquid and in polarized
paramagnetic Fermi liquid coincide with those described at the end of the
Section II.

\section{Conclusion}

In conclusion we stress once again that the transverse spin wave dispersion
in polarized paramagnetic Fermi liquid as well in a Fermi liquid with
spontaneous magnetization is found to be attenuating.  The spin wave
frequency is proportional to $k^{2}¥$ with complex diffusion coefficient
such that the damping at $T=0$ has a finite value proportional to the
scattering rate of quasiparticles.  This behavior of polarized
paramagnetic or feromagnetic Fermi liquid contrasts with the behavior
of Heisenberg ferromagnet in hydrodynamic regime \cite{Hal} where the
transverse spin wave attenuation appears in terms proportional $k^{4}¥$.

\section{Acknowledgements}

It is my pleasure to express the gratitude to G.Jackely, A.Meyerovich,
I.Fomin, W.Mullin and A.Huxley for the numerous stimulating
discussions.  I would also like to thank G.Vermeulen, E.Kats and
A.Vedyaev for the interest to the subject, and K.Kikoin for valuable
comments.


\begin{thebibliography}{99}

\bibitem{Sil} V.P.Silin, Zh.  Eksp.Teor.Fiz.  {\bf 33}, 1227 (1957)
[Sov.  Phys.JETP {\bf 6}, 945 (1958)]

\bibitem{Leg} A.J.Leggett, J.Phys.C {\bf 3}, 448 (1970).


\bibitem{JM} J.W.Jeon and W.J.Mullin, Phys.Rev.Lett.  {\bf 62}, 2691
(1989); W.J.Mullin and J.W.Jeon, Journ.  of Low Temp.Phys.  {\bf 88},
433 (1992).

\bibitem{Me}A.E.Meyerovich and K.A.Musaelian, Journ.  of Low
Temp.Phys.  {\bf 89}, 781 (1992); Phys.Rev.B {\bf 47}, 2897 (1993);
Journ.  of Low Temp.Phys.  {\bf 94}, 249 (1994).

\bibitem{Mu}A.E.Meyerovich and K.A.Musaelian, Journ.  of Low Temp.Phys. 
{\bf 89}, 781 (1992)

\bibitem {Gol} D.I.Golosov and A.E.Ruckenstein, Phys.Rev.Lett.  {\bf
74}, 1613 (1995); Journ.  of Low Temp.Phys.  {\bf 112}, 265 (1998).

\bibitem{Mat}A.E.Meyerovich and K.A.Musaelian, Phys.Rev.Lett.  {\bf
72}, 1710 (1994).

\bibitem{Min}V.P.Mineev, Phys.  Rev.  B {\bf 69}, 144429 (2004).

\bibitem{Akim}
H.Akimoto, D.Candela, J.S.Xia, W.J.Mullin, E.D.Adams and N.S.Sullivan,
Phys.Rev.Lett.  {\bf 90}, 105301 (2003). 

\bibitem{Ver} G.Vermeulen and A.Roni, Phys.Rev.Lett.{\bf 86}, 248 (2001).

\bibitem{Fom} I.A.Fomin, Pis'ma Zh.  Eksp.Teor.Fiz {\bf 65}, 717 (1997) [JETP
Lett.{\bf 65},749 (1997)].

\bibitem{Maki}K.Maki, Phys.Rev.B {\bf 11}, 4264 (1976).

\bibitem{Landau}L.D.Landau , Zh.  Eksp.Teor.Fiz.  {\bf 35}, 97 (1958)
[Sov.Phys.JETP {\bf 8}, 70 (1959)].

\bibitem{AbrDz}A.A.Abrikosov and I.E.Dzyaloshinskii, Zh.  Eksp.Teor.Fiz
{\bf 35}, 771 (1958) [Sov.Phys.JETP {\bf 8}, 535 (1958)].  This paper
is also reproduced as Appendix to the book by A.A.Abrikosov {\it
Fundamentals of the theory of metals}, Elsevier Science Publisher;
B.V. (1988)

\bibitem{Kondr}P.S.Kondratenko, Zh.  Eksp.Teor.Fiz.  {\bf 46}, 1438 (1964)
[Sov.Phys.JETP {\bf 19}, 972 (1964)].

\bibitem{DzKondr}I.E.Dzyaloshinskii, P.S.Kondratenko, Zh.  Eksp.Teor.Fiz. 
{\bf 70}, 1987 (1976) [Sov.Phys.JETP {\bf 43}, 1036 (1976)].

\bibitem{Rod} A. Rodrigues, G.Vermeulen, Journ.  Low Temp.Phys.  {\bf 108},
103 (1997).

\bibitem{Can} D.Candela, N.Matsuhara, D.S.Sherill and D.O.Edwards,
Journ.  Low Temp.  Phys.  {\bf 63}, 369 (1986).

\bibitem{Mor} T. Moriya "Spin fluctuations in itinerant elecron magnetism",
Springer-Verlag, Berlin, 1985.


\bibitem{Hal} B.I.Halperin and P.C.Hohenberg, Phys.Rev.  {\bf 188}, 898
(1969).

\bibitem{LifPit}E.M.Lifshits and L.P.Pitaevskii, "Statistical
Physics'', Part 2 (Pergamon Press, Oxford, 1980)

\bibitem{Ward} This property should be fullfilled automatically (sort of
Ward identity) in a theory with self-consistent treatment of collisions.


\bibitem{Her}C.Herring "Exchange Interactions among Itinerant Electrons"
 Chapter XIV, pp.345-385, in "Magnetism" v.IV, edited by G.T.Rado and
 H.Suhl, Academic Press, NY and London, 1966.



 




\end{thebibliography}
\end{document}